\documentclass[
twocolumn,
hf,
]{ceurart}

\usepackage{listings}
\usepackage{multirow}
\usepackage{tabularx}
\usepackage{enumitem}
\usepackage{graphicx}
\usepackage{subcaption}
\usepackage{placeins}
\usepackage{graphics}
\usepackage{caption}
\graphicspath{{figures/}}

\begin{document}
\copyrightyear{2024}
\copyrightclause{Copyright for this paper by its authors.
  Use permitted under Creative Commons License Attribution 4.0
  International (CC BY 4.0).}

%%%%%%%%%%%%%%%%%%%%%%%%%%%%%%%%%%%%%%%%%%%%%%%%%%%%%%%%%%%%%%%%v
\conference{47th German Conference on Artificial Intelligence, 2nd Workshop on Public Interest AI, Monday, 23 September 2024}
%%%%%%%%%%%%%%%%%%%%%%%%%%%%%%%%%%%%%%%%%%%%%%%%%%%%%%%%%%%%%%%%
\title{Data-driven Modeling of Combined Sewer Systems for Urban Sustainability: An Empirical Evaluation}

%%%%%%%%%%%%%%%%%%%%%%%%%%%%%%%%%%%%%%%%%%%%%%%%%%%%%%%%%%%%%%%%

\author[1]{Vipin Singh}[email=s91001@bht-berlin.de]
\author[2]{Tianheng Ling}[email=tianheng.ling@uni-due.de]
\author[1]{Teodor Chiaburu}[email=chiaburu.teodor@bht-berlin.de]
\author[1,3]{Felix Biessmann}[email=felix.biessmann@bht-berlin.de]
\address[1]{Berlin University of Applied Sciences and Technology, Luxemburger Str. 10, 13353 Berlin, Germany}
\address[2]{University of Duisburg-Essen, Bismarckstraße 90, 47057 Duisburg, Germany}
\address[3]{Einstein Center for Digital Future, Wilhelmstraße 67, 10117 Berlin, Germany}
%%%%%%%%%%%%%%%%%%%%%%%%%%%%%%%%%%%%%%%%%%%%%%%%%%%%%%%%%%%%%%%%v
\begin{abstract}
Climate change poses complex challenges, with extreme weather events becoming increasingly frequent and difficult to model. Examples include the dynamics of Combined Sewer Systems (CSS). Overburdened CSS during heavy rainfall will overflow untreated wastewater into surface water bodies.
Classical approaches to modeling the impact of extreme rainfall events rely on physical simulations, which are particularly challenging to create for large urban infrastructures. Deep Learning (DL) models offer a cost-effective alternative for modeling the complex dynamics of sewer systems. 
In this study, we present a comprehensive empirical evaluation of several state-of-the-art DL time series models for predicting sewer system dynamics in a large urban infrastructure, utilizing three years of measurement data. We especially investigate the potential of DL models to maintain predictive precision during network outages by comparing global models, which have access to all variables within the sewer system, and local models, which are limited to data from a restricted set of local sensors.
Our findings demonstrate that DL models can accurately predict the dynamics of sewer system load, even under network outage conditions. These results suggest that DL models can effectively aid in balancing the load redistribution in CSS, thereby enhancing the sustainability and resilience of urban infrastructures. 
\end{abstract}
%%%%%%%%%%%%%%%%%%%%%%%%%%%%%%%%%%%%%%%%%%%%%%%%%%%%%%%%%%%%%%%%
\maketitle
%%%%%%%%%%%%%%%%%%%%%%%%%%%%%%%%%%%%%%%%%%%%%%%%%%%%%%%%%%%%%%%%

\section{Introduction and Related Work}
\label{sec:introduction}

% Background and Chanllegnes
Climate change has increased the frequency and intensity of extreme weather events~\cite{bolan2023impacts}, which pose significant challenges to urban infrastructure and environmental management~\cite{wilbanks2013climate}. Managing Combined Sewer Systems (CSS) becomes particularly difficult~\cite{wang2013consequential}. Heavy rainfall can overwhelm the capacity of these systems, leading to overflows that release untreated sewage into rivers and lakes~\cite{botturi2021combined}. This contamination compromises water quality and poses direct risks to human health~\cite{sonone2020water}.

\begin{figure*}[!htb]
    \centering
    \includegraphics[width=0.7\textwidth]{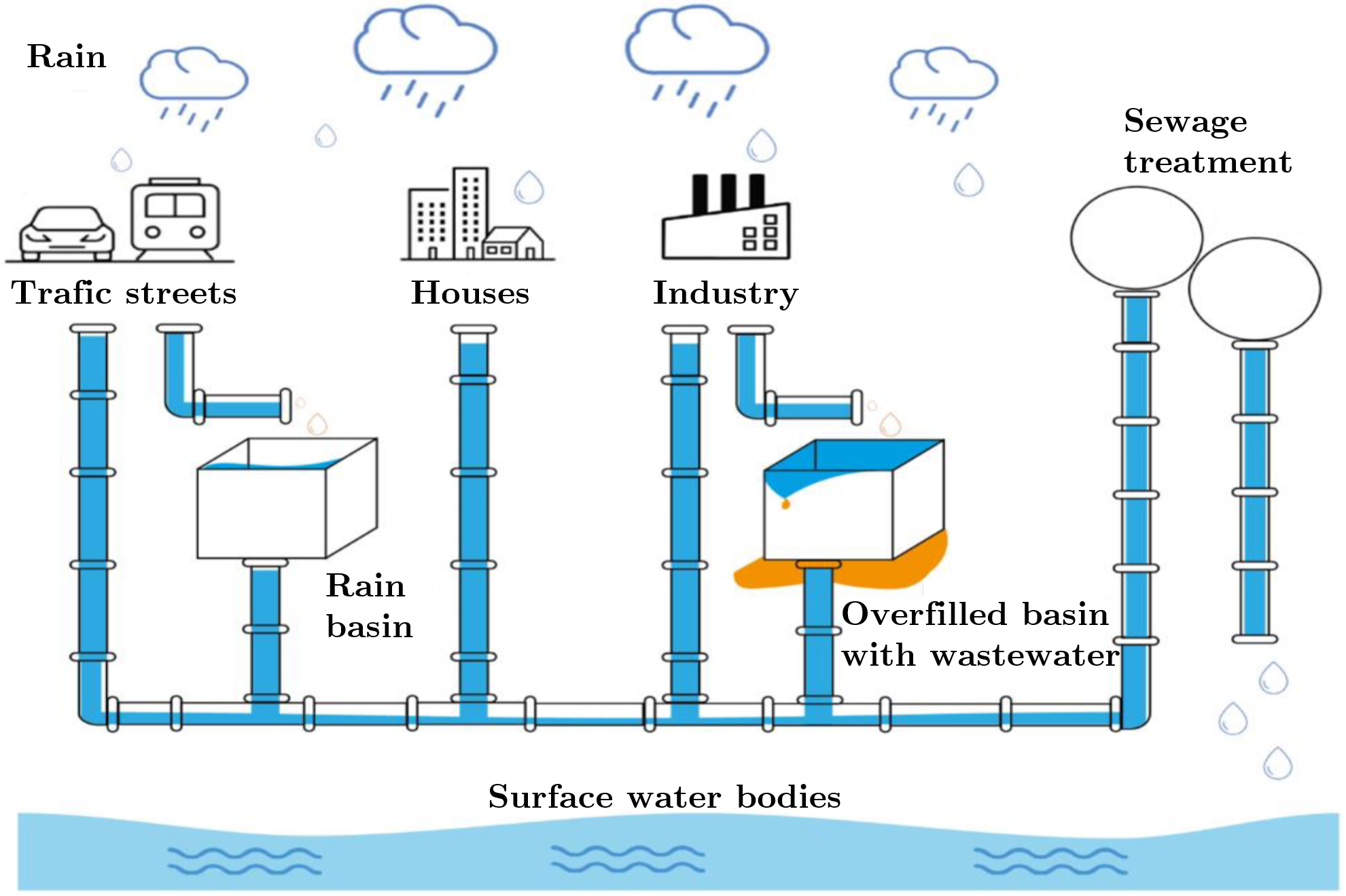}
    \caption{CSS with overflow basins collect rainwater and wastewater into multiple basins. During heavy rainfall, these basins may exceed capacity, leading to the untreated mixture overflow into the environment~\cite{chiaburu2024interpretable}.}
    \label{fig:CSS}
\end{figure*}

Many urban areas that utilize CSS have implemented overflow basins to mitigate this risk~\cite{botturi2021combined}, as shown in Figure~\ref{fig:CSS}. However, there remains a significant gap in understanding the dynamics of water levels in these overflow basins.
Traditional methods for modeling the dynamics of sewer systems rely on physical simulations~\cite{schutze2002modelling}. These systems are challenging to apply to large urban infrastructures as they require domain expertise and detailed data on the system components, which is often unavailable or imposes significant financial costs.
Improving the forecasting of these water levels can significantly enhance real-time flow control and inform maintenance and extension planning for sewage overflows.

% General Solution
Data-driven approaches, such as Deep Learning (DL) models, particularly time-series models, offer a promising alternative for modeling sewer system dynamics. Any target variable can be modeled with a combination of variables of the sewer system and exogenous variables, such as rainfall, without an explicit model of the sewer system, as it would be required for classical hydrological systems. These models enable accurate and flexible modeling of sewage treatment facilities to proactively manage and redistribute the load, thus preventing overflows and mitigating their impacts~\cite{saddiqi2023smart}. Such predictive capabilities are crucial for timely interventions and informed decision-making in urban water management.

% Public interest AI part
Transforming urban infrastructure to address the challenges of extreme weather on urban sewer systems requires cost-efficient modeling. Here, we investigate the potential of data-driven methods and demonstrate the potential of DL-based approaches to foster a more equitable and sustainable urban environment.
Our data-driven models enhance the predictive capabilities of sewer system management, reducing the likelihood of untreated sewage overflows while keeping financial costs lower than traditional methods, thereby making proactive management accessible to lower-income and marginalized communities.
These models also protect public health by minimizing water contamination, particularly in vulnerable populations, and ensure cleaner water bodies.
Additionally, this solution enables cities to optimize resource allocation and emergency responses, reducing environmental damage and long-term economic costs.
It underscores the importance of deploying advanced technologies in ways that prioritize the well-being of all community members, ensuring that technological advancements contribute to the common good and support populations in the face of climate change, aligning with the principles of public interest AI~\cite{zueger2023piai}.

It is worth mentioning that there are approaches aiming at combining classical physics-based simulations with the flexibility of DL methods \cite{raissi2018hfm}. The challenge with these combinations is that the model architectures require the same detailed knowledge about the sewer system as traditional methods. In addition, modeling the mixed viscosity of wastewater imposes significant complexity on the physical models. Here we focus on a data-driven approach that learns the system dynamics from measurements and is thus readily applicable without building cost-intense digital twins reflecting the physical properties of the sewer system.

% -----------------------------------------
% -----------------------------------------

% Our Solution and Contributions
In this work, we evaluate the performance of various advanced time series models for forecasting the water levels in CSS overflow basins. We explore and compare two approaches to time series forecasting: (1) a global model approach, which incorporates a number of exogenous variables, such as rainfall data, and (2) a local model approach, which relies solely on historical water level data. Our analysis aims to identify the most effective models and approaches for practical application in sewage overflow management.
Our study extends prior work on time series models for wastewater modeling~\cite{chiaburu2024interpretable} with the following main contributions:

\begin{itemize}
    \item \textbf{Comprehensive Model Evaluation}: We systematically evaluated multiple state-of-the-art time series models based on three years of real-world data. The Long Short-Term Memory (LSTM)~\cite{hochreiter1997long} and Temporal Fusion Transformer (TFT)~\cite{lim2021temporal} models, in particular, showed superior performance, with LSTM achieving the lowest Mean Squared Error (MSE) and TFT providing robust and consistent predictions across various conditions.

    \item \textbf{Global vs. Local Model Comparison}: We compared global and local model approaches, highlighting their strengths and limitations in sewage overflow forecasting. Our findings indicate that global models generally outperform local models in terms of MSE. However, local models are advantageous in scenarios where exogenous data is unavailable, offering a computationally efficient alternative.
\end{itemize}

%%%%%%%%%%%%%%%%%%%%%%%%%%%%%%%%%%%%%%%%%%%%%%%%%%%%%%%%%%%%%%%%
\section{Data and Preprocessing}
\label{sec:data}

In our study, we utilized data\footnote{Data cannot be made publicly available. Readers can contact the corresponding author for details.} provided by \emph{Wirtschaftsbetriebe Duisburg}\footnote{\url{https://www.wb-duisburg.de}}. The dataset comprises time series sensory data, including water levels in rain basins and water tanks, energy consumption of pumps, and rainfall amounts. The sensor data were collected from six locations in Duisburg's Vierlinden district, covering three years, from January 1, 2021, at 00:00 AM until January 1, 2024, at 00:00 AM. The recording intervals are irregular as the sensors were read out event-based, with sensor update intervals ranging from 1 second to 1 hour. 

To standardize the data, we resampled it by calculating the mean values closest to each full-hour mark. This resampling procedure resulted in a total of 26,280 data points, with each data point comprising 35 features derived from the sensory data across the different locations. For missing values in the rainfall measurements, we utilized data from the nearest weather station of the \emph{Deutsche Wetterdienst}\footnote{\url{https://www.dwd.de/DE/Home/home_node.html}}, specifically the station in Duisburg-Baerl, located 4.5 km from the sewage treatment plant that recorded the rainfall. For the other features, linear interpolation was employed to fill the missing values. Additionally, an indicator column was added for each feature with missing values to denote whether the corresponding value was interpolated.

%%%%%%%%%%%%%%%%%%%%%%%%%%%%%%%%%%%%%%%%%%%%%%%%%%%%%%%%%%%%%%%%
\section{Methodology}
\label{sec:methodology}

This section details the time series models employed in this study, including the selection, implementation, and comparisons of global and local model approaches.

%%%%%%%%%%%%%%%%%%%%%%%%%%%%%%%%%%%%
\subsection{Neural Network Architectures}

For our empirical evaluation, we selected six state-of-the-art neural time series models. While classical regression models, such as tree-based methods, can be effective for time series data, the state of the art in water modeling increasingly relies on neural network models \cite{MAIER2000101}. We thus focus on these models based on their effectiveness and versatility in forecasting tasks. The selected models are:
\begin{itemize}
    \item LSTM~\cite{hochreiter1997long}: LSTM networks are well-suited for capturing long-term dependencies in sequential data, making them ideal for time series forecasting.
    \item DeepAR~\cite{salinas2020deepar}: This probabilistic forecasting model leverages autoregressive recurrent networks, providing robust predictions with uncertainty estimates.
    \item Neural Hierarchical Interpolation for Time Series Forecasting (N-HiTS)~\cite{challu2023nhits}: As a neural hierarchical time series model, N-HiTS excels in capturing complex temporal patterns.
    \item Transformer~\cite{wen2022transformers}: Originally designed for nature language processing~\cite{vaswani2017attention}, Transformers use attention mechanism to effectively capture relationships across different time steps~\cite{wen2022transformers}.
    \item Temporal Convolutional Network (TCN)~\cite{bai2018empirical}: TCNs can model long-range dependencies in time series data while being computationally efficient.
    \item TFT~\cite{lim2021temporal}: TFT combines the strengths of LSTM and attention mechanisms to provide interpretable and accurate forecasts.
\end{itemize}

These models were implemented using the \emph{Darts}\footnote{\url{https://unit8co.github.io/darts/}} Python library, which offers a user-friendly interface for time-series forecasting. \emph{PyTorch}\footnote{\url{https://pytorch.org}} was used as the supporting framework.

%%%%%%%%%%%%%%%%%%%%%%%%%%%%%%%%%%%%
\subsection{Global vs Local Model Approach}

Our study considers two approaches to time series forecasting, each motivated by distinct real-world scenarios. 

\begin{enumerate}
    \item \textbf{Global Model Approach}: This approach corresponds to the scenario of normal CSS operation, where all sensors are fully operational, and all data can be transmitted reliably over the network. In this case, all available data, including exogenous variables such as rainfall data, are integrated into a single model for forecasting the relevant target variable. This approach, referred to as the \textit{global model}, allows the models to leverage additional contextual information to improve forecasting precision.
    
    \item \textbf{Local Model Approach}: This approach is designed for scenarios where not all data is available. Such circumstances can arise when sensors are damaged or network connections are unstable due to extreme weather events, or security incidents. To mimic these cases, predictions are made using only the historical data from the specific sensor in question, without access to additional contextual information. This approach is referred to as the \textit{local model}. It is intended for future deployment on edge devices, enabling localized and resilient forecasting capabilities~\cite{ling2024flowprecision}.
\end{enumerate}

Overall, the global model approach leverages extensive data to enhance forecasting precision, whereas the local model approach ensures robustness and adaptability in environments with limited data availability. By evaluating both approaches, we aim to provide a comprehensive solution for diverse operational scenarios in urban wastewater management.

%%%%%%%%%%%%%%%%%%%%%%%%%%%%%%%%%%%%%%%%%%%%%%%%%%%%%%%%%%%%%%%%
\section{Experiments}
\label{sec:experiments}

This section introduces the experimental settings, including data splits, model development, and error metrics.

%%%%%%%%%%%%%%%%%%%%%%%%%%%%%%%
\subsection{Datasets}
The dataset was divided into training, validation, and testing sets. The first two years of data were used for training and validation, while the last year was reserved for testing. Within the initial two years, 80\% of the data was allocated for training and 20\% for validation. Given the sequential nature of time series data, the data were not shuffled, and the split was performed in chronological order to maintain temporal dependencies. To ensure consistency across features, standard scaling was applied using default parameters ($\mu = 0$, $\sigma^2 = 1$) prior to model training.

After fine-tuning (see \autoref{app:hpo}), we determined that a 72-hour input sequence was optimal for forecasting a 12-hour prediction sequence. This prediction sequence was determined together with domain experts to meet operational requirements.

%%%%%%%%%%%%%%%%%%%%%%%%%%%%%%%
\subsection{Model Development}

To prevent overfitting, early stopping with patience of 10 epochs was adopted. All models were optimized using the \emph{Adam} optimizer~\cite{kingma2014adam} with 32-bit floating point precision. Training sessions were conducted on an NVIDIA A100 with 40GB VRAM, utilizing CUDA 12.2 and Python 3.10. Hyperparameter optimization was performed using the Tree-structured Parzen Estimator algorithm provided by the \emph{Optuna} library\footnote{\url{https://optuna.org/}}. Each model had a training budget of 600 trials, with each trial consisting of 100 epochs. Further details on the hyperparameter optimization process are provided in \autoref{app:hpo}. The best hyperparameter configuration for each model was then evaluated using 100 different random weight initializations to ensure robust comparisons.

%%%%%%%%%%%%%%%%%%%%%%%%%%%%%%%
\subsection{Error Metrics}

We compared various error metrics well established in the field of time series forecasting~\cite{hewamalage2021look}, including MSE and  Mean Absolute Percentage Error (MAPE). For model training, MSE was selected as the loss function, except for the probabilistic model DeepAR, which utilized the negative log-likelihood. While both metrics have limitations, they are among the most often used metrics to evaluate regression tasks. Another advantage of these metrics is that they can be easily interpreted by domain experts operating the sewer system. We opted for these established metrics for comparability, reproducibility, and practical applicability. 

%%%%%%%%%%%%%%%%%%%%%%%%%%%%%%%%%%%%%%%%%%%%%%%%%%%%%%%%%
\section{Results}

\begin{figure}[!htb]
    \centering
    \begin{subfigure}[b]{\linewidth}
        \centering
        \includegraphics[width=\textwidth]{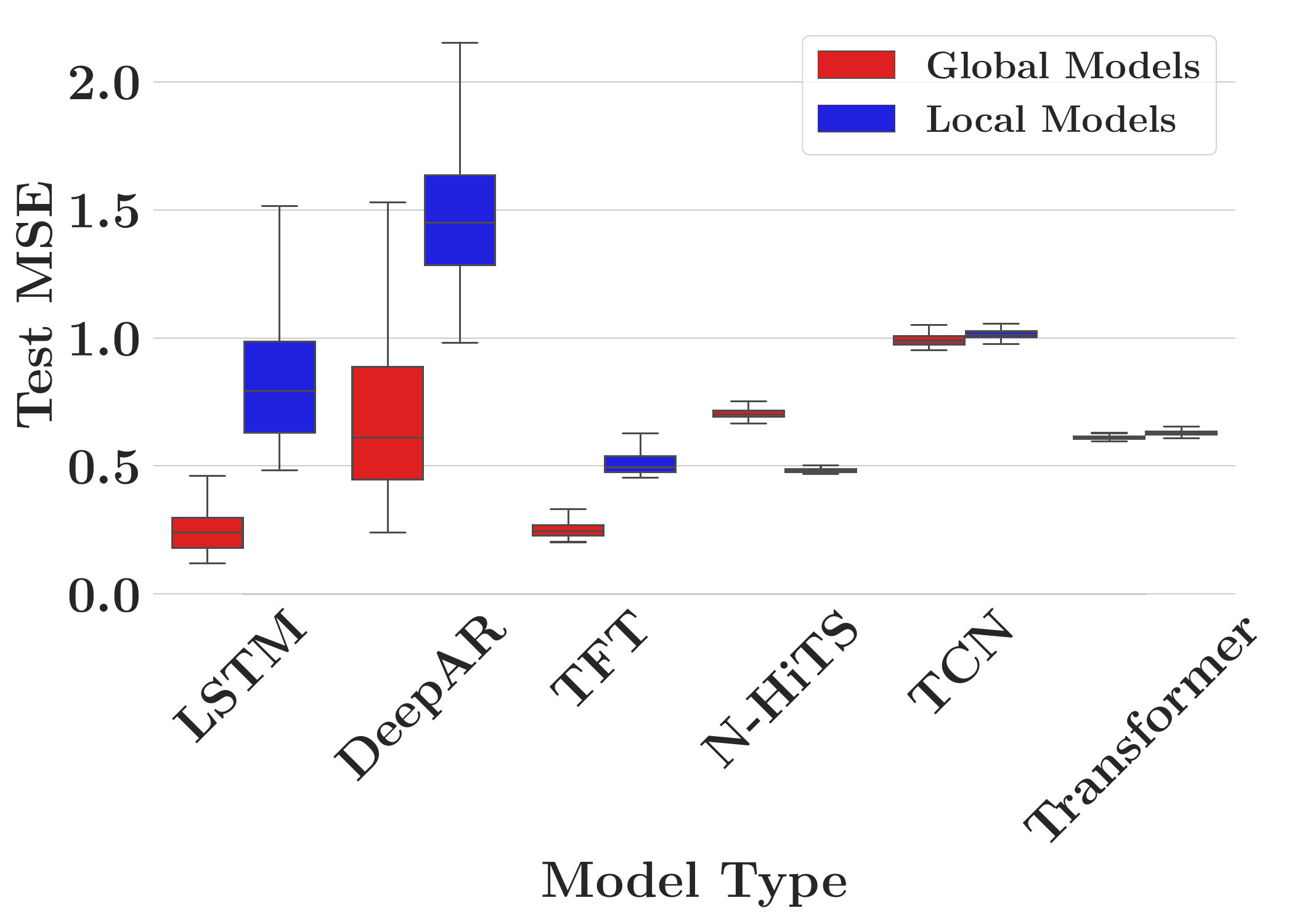}
    \end{subfigure}
    \hfill
    \begin{subfigure}[b]{\linewidth}
        \centering
        \includegraphics[width=\textwidth]{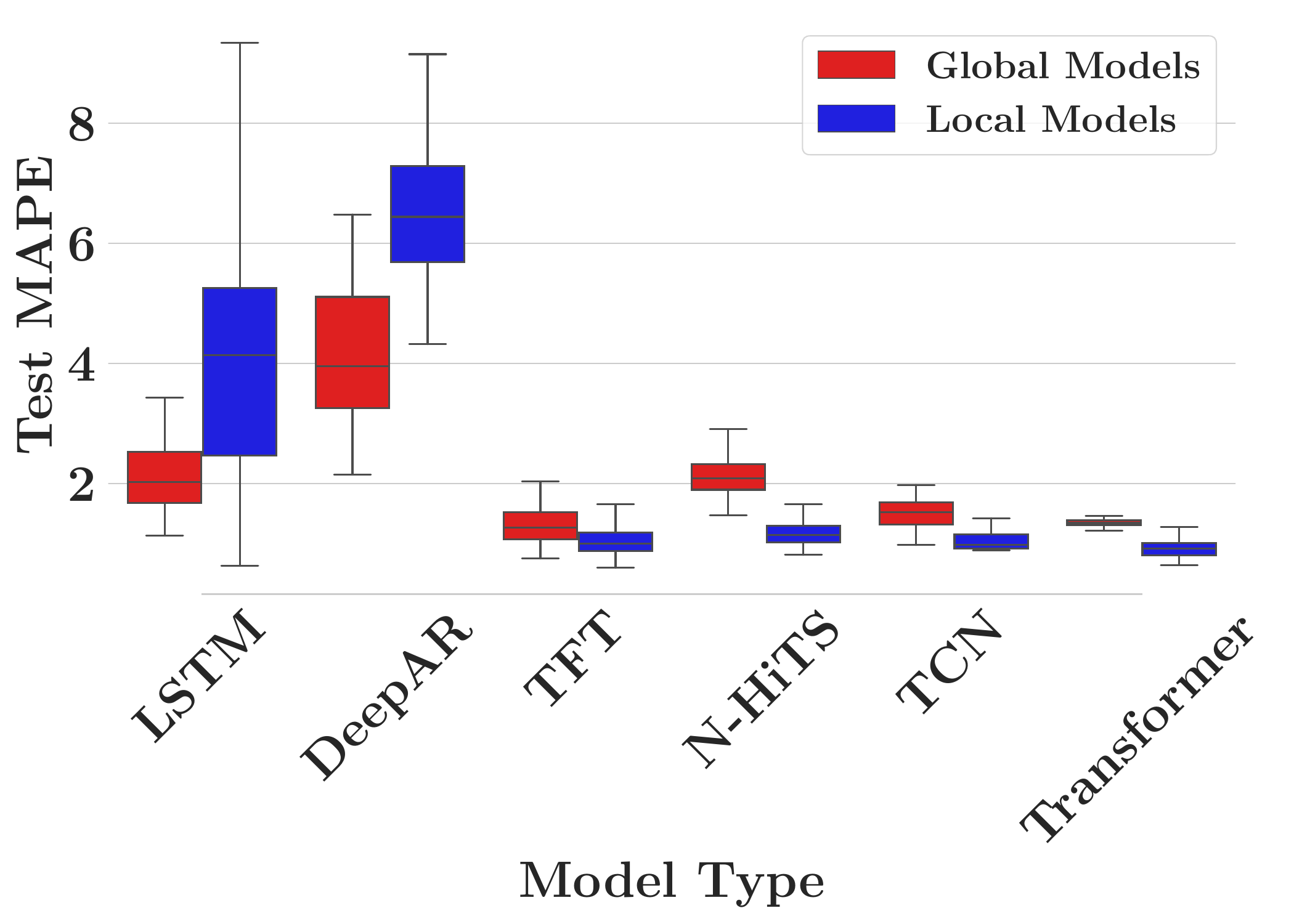}
    \end{subfigure}
    \caption{Global models often outperform the local models. The distribution of test MSE (upper) and MAPE (lower) across model types and approaches, based on multiple training runs with different random parameter initializations, show that the lowest median MSE values were obtained with LSTM and TFT models. The lowest median MAPE values were obtained with even simpler models such as TCN.}
    \label{fig:error_bars}
\end{figure}

\autoref{fig:error_bars} shows the distribution of test MSE and MAPE across the 100 random weight initializations for each model type and approach. It is evident in both figures that the LSTM and DeepAR models have a larger spread in the results compared to the other models.
% ------------------------------------
% ------------------------------------
% ------------------------------------
\begin{table*}[!htb]
    \centering
    \caption{Considering MSE, inference time, and model size, the LSTM and TFT models stand out among global and local models. This comparison includes MSE, mean inference time per sample, and effective model size for both global and local approaches. The best values are highlighted in green.}
    \resizebox{\textwidth}{!}{
    \begin{tabular}{|c|cc|cc|cc|cc|cc|}
        \toprule
        \multirow{3}{*}{Models} & \multicolumn{6}{c|}{MSE}                                                                                                                                  & \multicolumn{2}{c|}{\begin{tabular}[c]{@{}c@{}}Inference {[}ms{]}\end{tabular}} & \multicolumn{2}{c|}{\begin{tabular}[c]{@{}c@{}}Size {[}MB{]}\end{tabular}} \\ \cline{2-11} 
                        & \multicolumn{2}{c|}{q=0.25}                              & \multicolumn{2}{c|}{q=0.5}                               & \multicolumn{2}{c|}{q=0.75}         & \multicolumn{1}{c|}{\multirow{2}{*}{Global}}          & \multirow{2}{*}{Local}          & \multicolumn{1}{c|}{\multirow{2}{*}{Global}}        & \multirow{2}{*}{Local}        \\ \cline{2-7}
                        & \multicolumn{1}{c|}{Global} & \multicolumn{1}{c|}{Local} & \multicolumn{1}{c|}{Global} & \multicolumn{1}{c|}{Local} & \multicolumn{1}{c|}{Global} & Local & \multicolumn{1}{c|}{}                                 &                                 & \multicolumn{1}{c|}{}                               &                               \\ \hline
        LSTM & \cellcolor{green}0.18 & 0.63 & \cellcolor{green}0.24 & 0.79 & 0.30 & 0.99 & \cellcolor{green}0.80 & \cellcolor{green}0.80 & 0.31 & 0.16 \\
        TFT & 0.23 & \cellcolor{green}0.48 & 0.25 & 0.50 & \cellcolor{green}0.27 & 0.54 & 3.08 & 1.40 & 11.25 & 6.22 \\
        DeepAR & 0.45 & 1.28 & 0.61 & 1.45 & 0.89 & 1.64 & 2.08 & 2.34 & 1.71 & 1.65 \\
        Transformer & 0.61 & 0.62 & 0.61 & 0.63 & 0.62 & 0.64 & 0.97 & 1.08 & 144.26 & 184.59 \\
        N-HiTS & 0.69 & \cellcolor{green}0.48 & 0.70 & \cellcolor{green}0.48 & 0.72 & \cellcolor{green}0.49 & 0.92 & 0.87 & 1138.41 & 27.10 \\
        TCN & 0.98 & 1.00 & 0.99 & 1.01 & 1.01 & 1.03 & 1.24 & 1.27 & \cellcolor{green}0.07 & \cellcolor{green}0.04 \\
        \bottomrule
    \end{tabular}
    }
    \label{tab:model_comparison}
\end{table*}
% ------------------------------------

In \autoref{tab:model_comparison}, we list the 0.25, 0.5, and 0.75 quantiles (q) of the MSE, along with the average measured wall clock runtime at single inference and the effective size of the trained model in Megabytes (MB). We observe that the inference times for global and local models are similar in most cases. This is expected, as the experiments were conducted on high-capacity hardware. It should be noted that these times were measured while executing the models in parallel on a single GPU.

% ------------------------------------
\begin{figure*}[!htbp]
    \centering
    % ------------------------------------
    \begin{subfigure}[b]{0.45\textwidth}
        \centering
        Global model (LSTM)
    \end{subfigure}
    \hfill
    \begin{subfigure}[b]{0.45\textwidth}
        \centering
        Local model (TFT)
    \end{subfigure}
% ------------------------------------
    \begin{subfigure}[b]{0.49\textwidth}
        \centering
        \includegraphics[trim=0 0 155px 57px, clip, width=\textwidth]{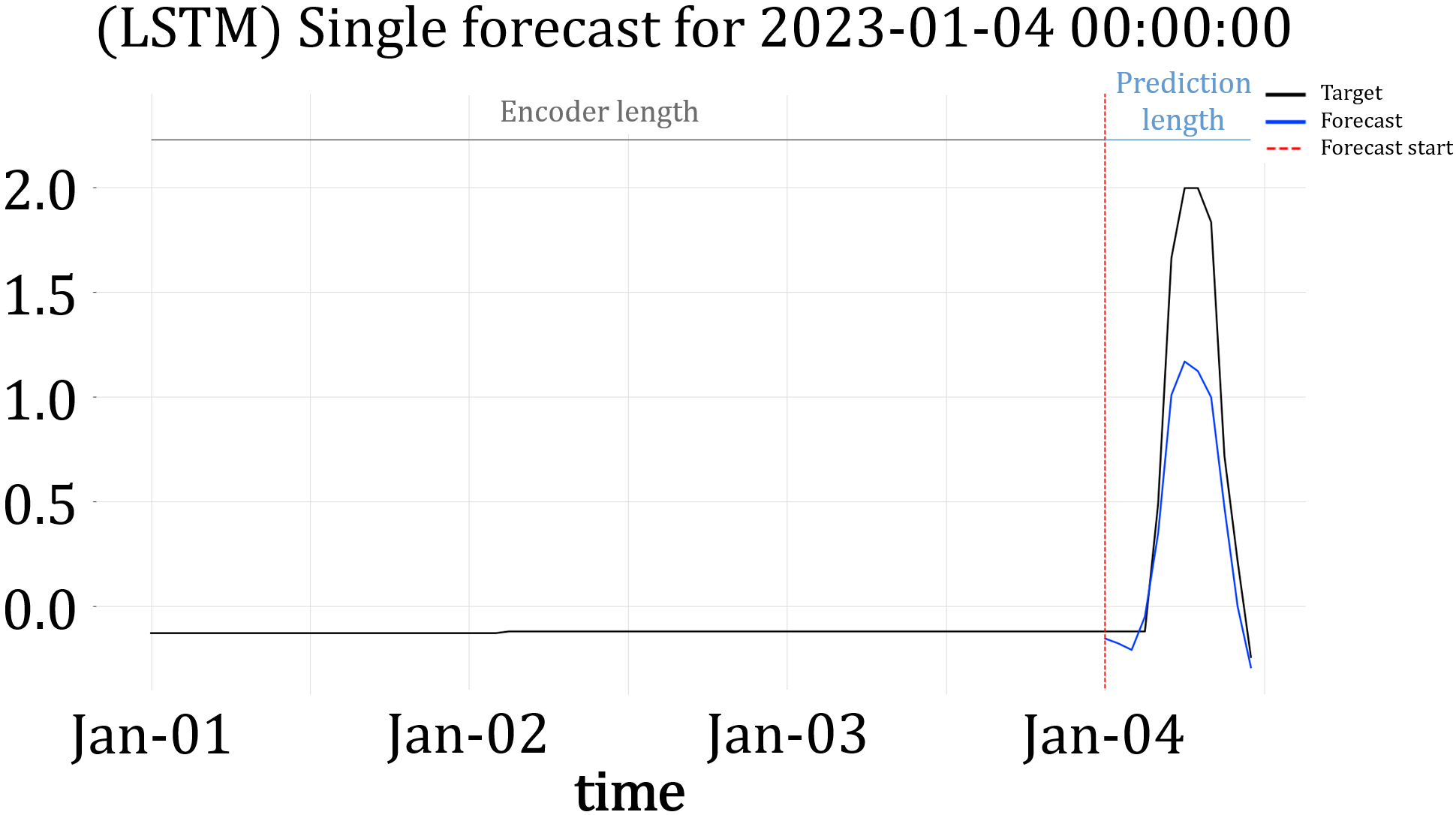}
        \caption{LSTM 12h Forecast for 2023-01-04 00:00:00}
        \label{fig:lstm_single1} 
    \end{subfigure}
    \hfill
    \vline
    \hfill
% ------------------------------------
    \begin{subfigure}[b]{0.49\textwidth}
        \centering
        \includegraphics[trim=0 0 155px 57px, clip, width=\textwidth]{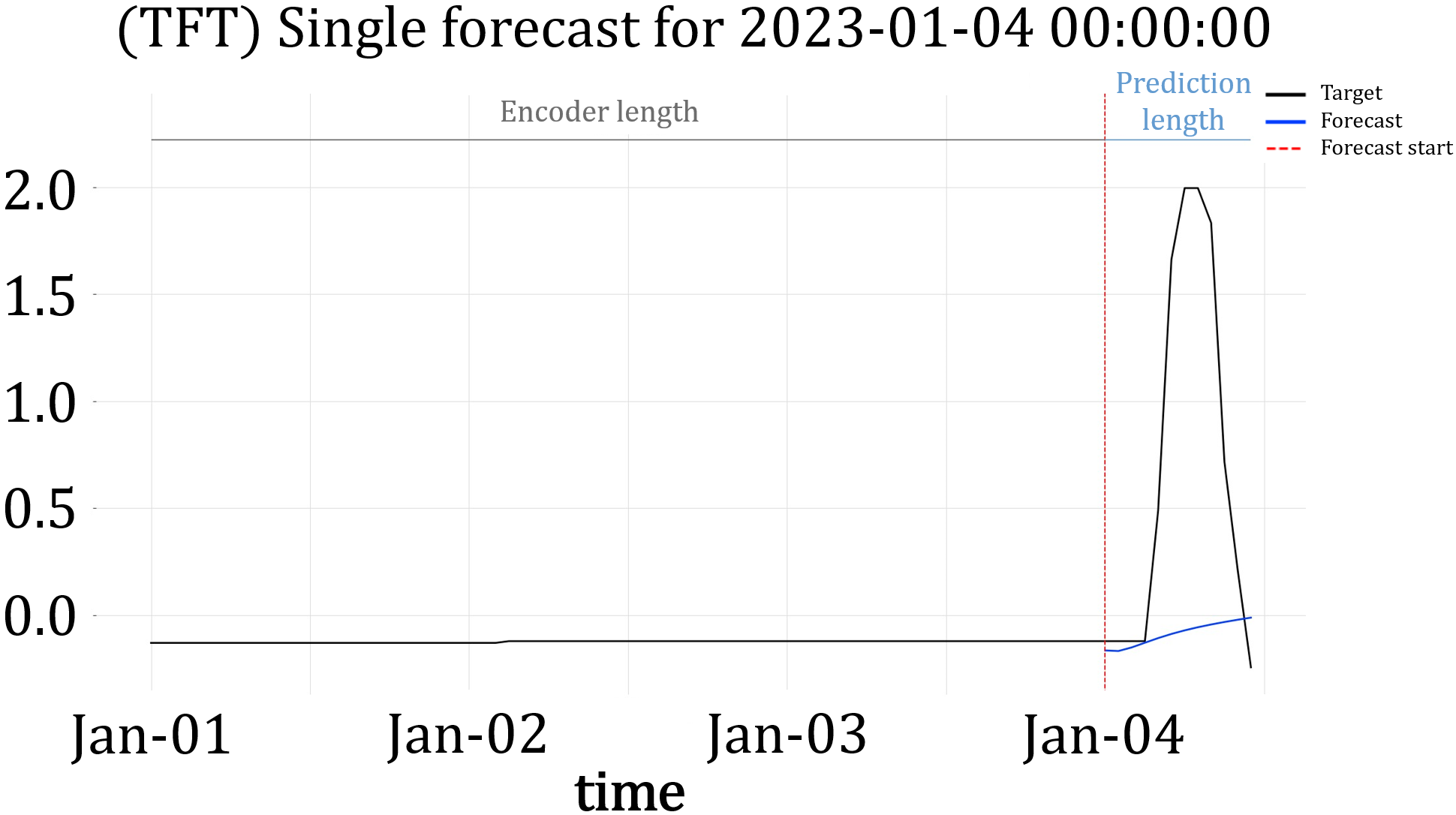}
        \caption{TFT 12h Forecast for 2023-01-04 00:00:00}
        \label{fig:tft_single1} 
    \end{subfigure}
% ------------------------------------
    \vspace{0.5cm} 
% ------------------------------------
    \begin{subfigure}[b]{0.49\textwidth}
        \centering
        \includegraphics[trim=0 0 155px 57px, clip, width=\textwidth]{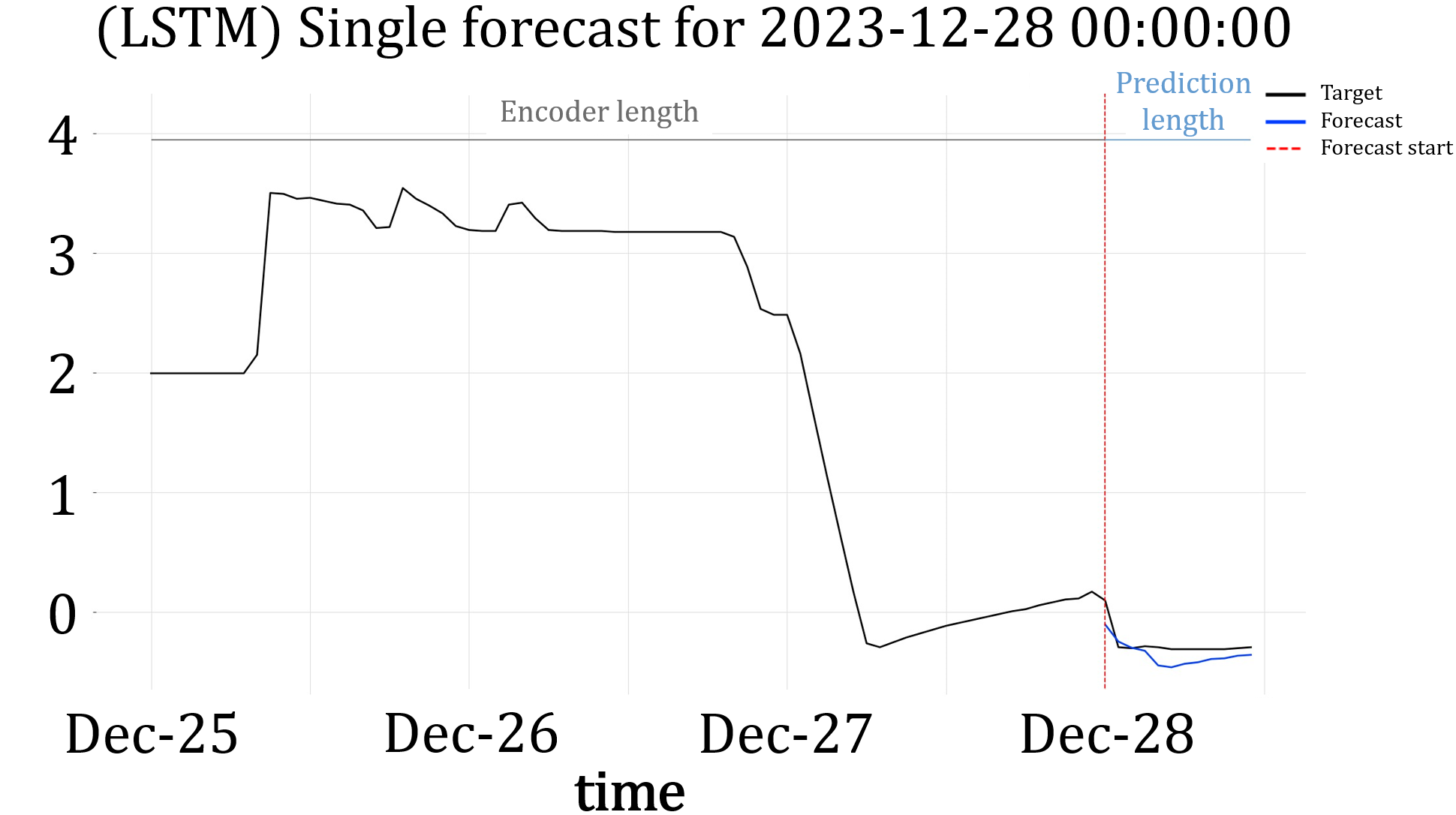}
        \caption{LSTM 12h Forecast for 2023-12-28 00:00:00}
        \label{fig:lstm_single2} 
    \end{subfigure}
    \hfill
    \vline
    \hfill
% ------------------------------------
    \begin{subfigure}[b]{0.49\textwidth}
        \centering
        \includegraphics[trim=0 0 155px 57px, clip, width=\textwidth]{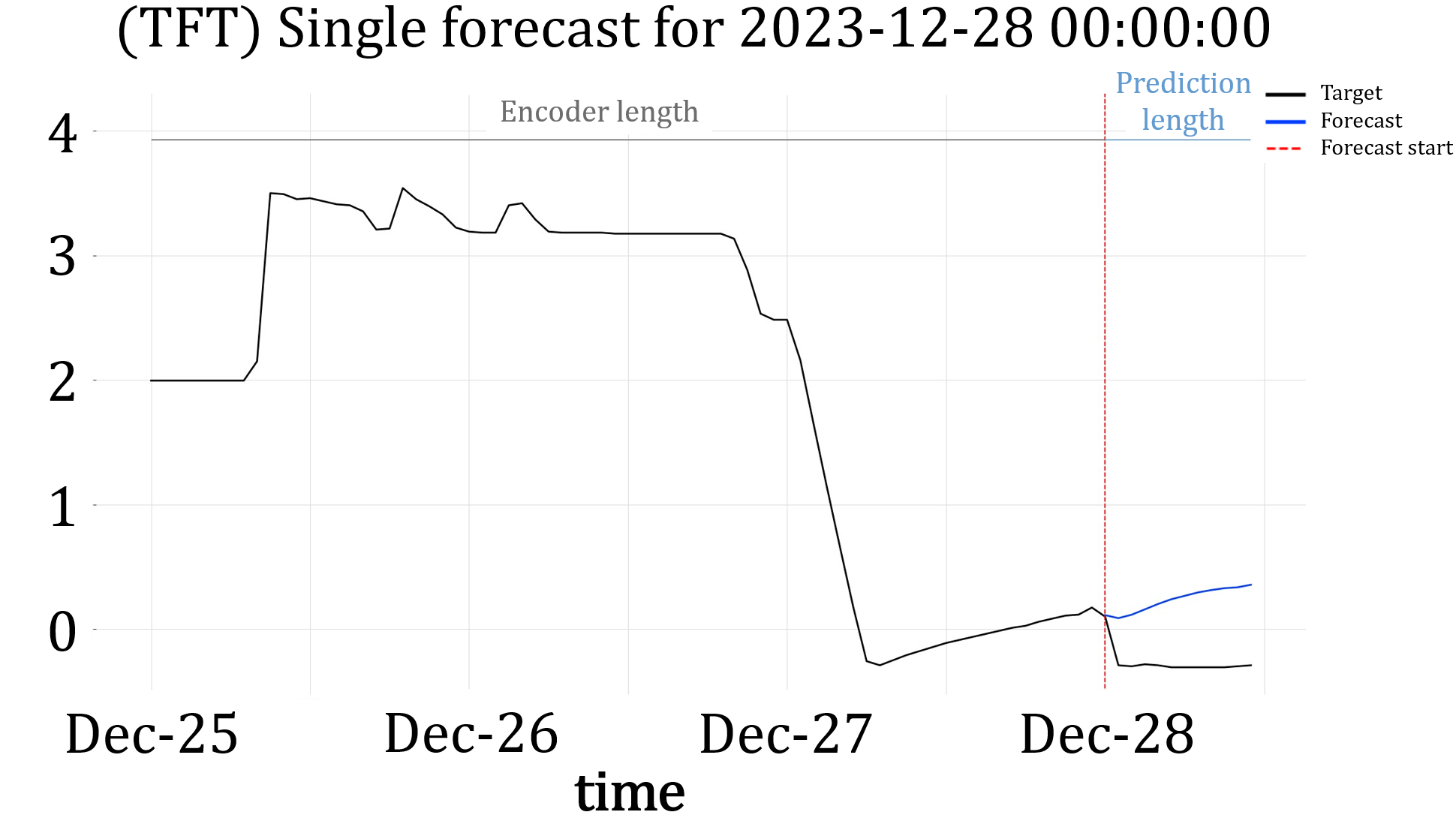}
        \caption{TFT 12h Forecast for 2023-12-28 00:00:00}
        \label{fig:tft_single2} 
    \end{subfigure}
    
    \begin{minipage}{\textwidth}
        \vspace{-20pt}
        \centering
        \includegraphics[width=0.5\textwidth]{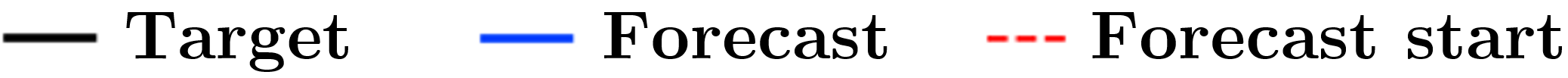}
        \label{fig:legend_lstm}
    \end{minipage}
    \vspace*{-15pt}
    \caption{The global LSTM model demonstrates better forecasting performance than the local TFT model. Forecasts from the global LSTM model (see Figure \ref{fig:lstm_single1} and \ref{fig:lstm_single2}) and local TFT model (see Figure \ref{fig:tft_single1} and \ref{fig:tft_single2}) were evaluated on two samples. The plots show the forecasts for the water level of the overflow basin.
    The dashed red line indicates the start of the forecast window.
    Samples before the dashed red line represent the input data for the model, spanning 72 hours, while the model produces a 12-hour forecast, shown in blue.}
    \label{fig:comparative_single_forecasts}
\end{figure*}
% ------------------------------------
Among the global models, the LSTM model has the best performance in terms of median MSE. Despite its larger spread in MSE, the LSTM model benefits from having the second lowest memory consumption and the fastest inference time. The TFT model shows the second-best performance among both global and local models, exhibiting low spread but having the highest measured inference time and relatively high memory consumption. In general, the LSTM and TFT models emerge as the most suitable models among the global and local approaches, respectively. However, we highlight the TCN model for its lowest memory consumption, which is five times less than the second-lowest model, the LSTM model. Another notable model is the N-HiTS, which achieves the lowest median MSE among the local models. Considering the memory consumption of the N-HiTS model, it is evident that it requires high-capacity resources for operation.

\autoref{fig:comparative_single_forecasts} shows representative examples of forecasts obtained from the LSTM and TFT models. The forecasts shown are from the experiments that achieved the lowest MSE for each model. The global model is represented by the LSTM, while the TFT represents the local model.

% ------------------------------------
\begin{figure*}[!htbp]
    \centering
    \begin{subfigure}[b]{0.49\textwidth}
        \centering
        \includegraphics[width=\textwidth]{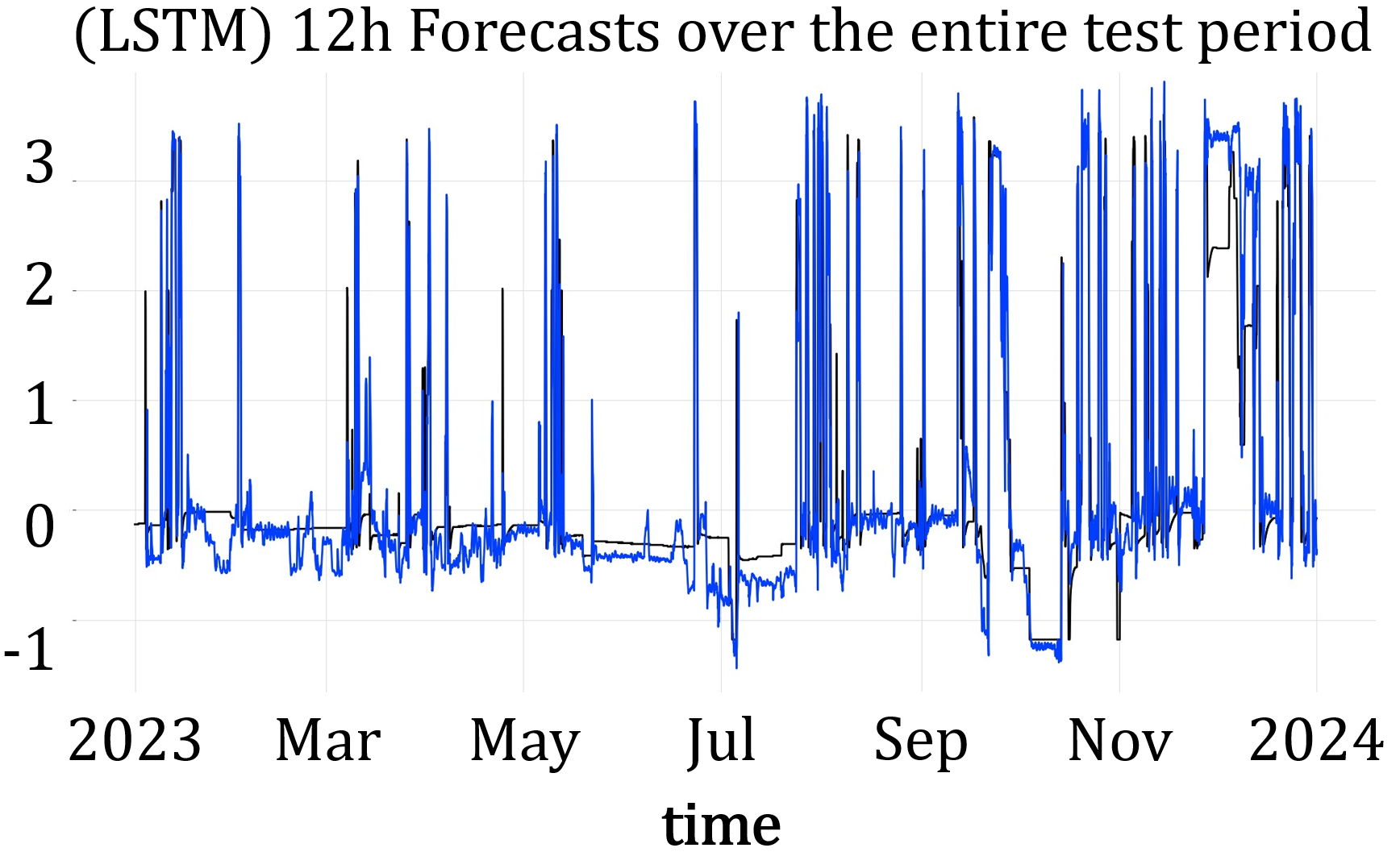}
    \label{fig:lstm_all_12h_forecasts}
    \end{subfigure}
    \hfill
    \begin{subfigure}[b]{0.49\textwidth}
        \centering
        \includegraphics[width=\textwidth]{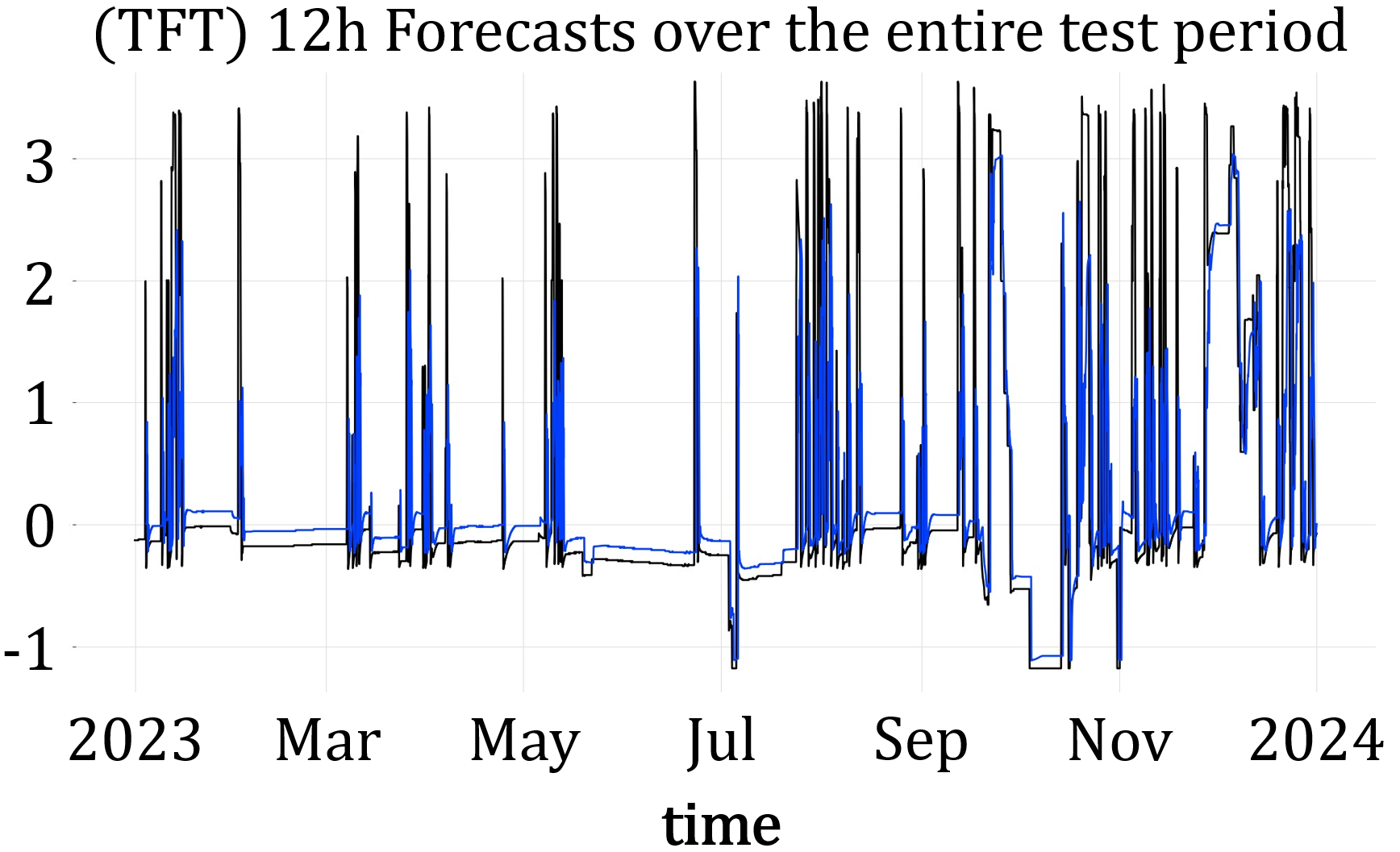}
    \label{fig:tft_all_12h_forecasts}
    \end{subfigure}
    
    \begin{minipage}{\textwidth}
        \vspace{-20pt}
        \centering
        \hspace{40pt}
        \includegraphics[width=0.5\textwidth]{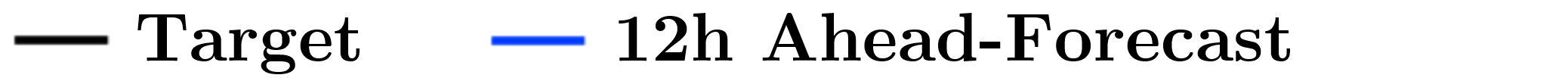}
        \label{fig:legend_all_forecasts}
    \end{minipage}
    \vspace*{-15pt}
    \caption{Global models are able to predict sudden changes much better than local models. 12-Hour Ahead Forecasts of the global LSTM (left) and local TFT (right) of Filling Levels Throughout 2023.}
    \label{fig:all_forecasts}
\end{figure*}
% ------------------------------------

In \autoref{fig:all_forecasts}, we present an exemplary forecast for a 12-hour horizon into the future. Although exhibiting considerable variability, we observe that the global LSTM model predicts spikes with a higher degree of precision. However, it shows considerable deviations around the mean values near zero. In contrast, the local TFT model struggles to predict sudden changes after longer periods of stagnancy.
Evidently, its forecast precision decreases as the forecast horizon extends.

%%%%%%%%%%%%%%%%%%%%%%%%%%%%%%%%%%%%%%%%%%%%%%%%%%%%%%%%%%%%%%%%
\section{Conclusion and Future Work}
\label{sec:conclusion_future}

Our results demonstrate that DL models can accurately predict the complex dynamics of wastewater levels in real-world scenarios. Global models, with full access to all sensor readings under normal operation without network outage, exhibit high forecast precision for wastewater levels in the overflow basin. This enhanced precision can significantly aid sewage treatment facilities in effectively redistributing the load of the CSS.

In contrast, local models perform worse in forecasting precision than global models. The reason could be the heavy concentration of target values around the mean. Due to sudden changes after longer periods of stability, the local models struggle with longer forecasting periods. However, local models can serve as a fallback in the event of a network interruption where exogenous variables become unavailable. Our results indicate that even when all network connections are lost, and only historical sensor readings of an individual sensor are available, adequate forecasts can still be made.
Additionally, due to their lower computational costs, it is worthwhile to explore the potential of deploying the local models on edge devices in the future.

\begin{acknowledgments}
The authors gratefully acknowledge the financial support provided by the Federal Ministry for Economic Affairs and Climate Action of Germany for the RIWWER project (Project number: 01MD22007H, 01MD22007C), the Einstein Center Digital Future in Berlin, and the German Research Foundation (Project number: 528483508 - FIP 12).
\end{acknowledgments}
%%%%%%%%%%%%%%%%%%%%%%%%%%%%%%%%%%%%%%%%%%%%%%%%%%%%%%%%%%%%%%%%
\bibliography{references}
\vfill
%%%%%%%%%%%%%%%%%%%%%%%%%%%%%%%%%%%%%%%%%%%%%%%%%%%%%%%%%%%%%%%%
\appendix
\section{Hyperparameter Optimization}
\label{app:hpo}

Hyperparameter optimization was performed using the Tree-structured Parzen Estimator algorithm provided in the \emph{Optuna} library\footnote{\url{https://optuna.org/}}. The optimization process was conducted in two iterations with a total of 600 trials:
\begin{enumerate}
    \item \textbf{Broad Search with 500 trials}: An extensive hyperparameter search space will be explored to identify potential optimal values.
    \item \textbf{Refined Search with 100 trials}: Based on the results of the broad search, a more focused and fine-grained search will be conducted around the best-performing hyperparameters.
\end{enumerate}

The hyperparameters optimized on the validation data include several hyperparameters shared by all models and some model-specific hyperparameters.

After the first iteration of fine-tuning, the input and prediction sequence lengths were set to 72 hours and 12 hours, respectively. Additionally, the batch size was set to 256, as this configuration worked well across all models. Below, we list the optimal hyperparameter values obtained after the second iteration for both the local and global models:

\begin{table}[width=\linewidth,pos=h]
    \centering
    \caption{Optimal Hyperparameter Settings for LSTM Models}
    \begin{tabular}{|l|l|l|}
        \hline 
        \multirow{2}{*}{Hyperparameters} & \multicolumn{2}{c|}{Optimal Values}             \\ \cline{2-3} 
                                 & \multicolumn{1}{c|}{Local Model} & Global Model \\ \hline
        % \textbf{Hyperparameter} & \textbf{Best value (local model)} & \textbf{Best value (global model)} \\ 
        learning\_rate & 0.0084 & 0.0228 \\
        batch\_size & 256 & 256 \\
        weight\_decay & 4.1575e-05 & 1.7515e-04 \\ 
        dropout & 0.2169 & 0.2895 \\
        hidden\_dim & 36 & 33 \\
        n\_rnn\_layers & 1 & 1 \\
        \hline
    \end{tabular}
    \label{tab:hyperparam_lstm}
\end{table}

\begin{table}[width=\linewidth,pos=h]
    \centering
    \caption{Optimal Hyperparameter Settings for DeepAR model}
    \begin{tabular}{|l|l|l|}
        \hline 
        \multirow{2}{*}{Hyperparameters} & \multicolumn{2}{c|}{Optimal Values}             \\ \cline{2-3} 
                                 & \multicolumn{1}{c|}{Local Model} & Global Model \\ \hline
        learning\_rate & 0.0460 & 0.0295 \\
        batch\_size & 256 & 256 \\
        weight\_decay & 2.004e-05 & 1.110e-05 \\ 
        dropout & 0.3142 & 0.3776 \\
        hidden\_dim & 128 & 106 \\
        n\_rnn\_layers & 1 & 1 \\
        \hline
    \end{tabular}
    \label{tab:hyperparam_deepar}
\end{table}

\begin{table}[width=\linewidth,pos=h]
    \centering
    \caption{Optimal Hyperparameter Settings for N-HiTS model}
    \begin{tabular}{|l|l|l|}
        \hline 
        \multirow{2}{*}{Hyperparameters} & \multicolumn{2}{c|}{Optimal Values}             \\ \cline{2-3} 
                                 & \multicolumn{1}{c|}{Local Model} & Global Model \\ \hline
        learning\_rate & 8.506e-05 & 2.611e-04 \\
        batch\_size & 256 & 256 \\
        weight\_decay & 4.014e-04 & 6.409e-03 \\ 
        dropout & 0.3027 & 0.5367 \\
        num\_stacks & 3 & 2 \\
        num\_blocks & 5 & 5 \\
        num\_layers & 1 & 1 \\
        layer\_widths & 1024 & 1024 \\
        \hline
    \end{tabular}
    \label{tab:hyperparam_nhits}
\end{table}

\begin{table}[width=\linewidth,pos=h]
    \centering
    \caption{Optimal Hyperparameter Settings for Transformer model}
    \begin{tabular}{|l|l|l|}
        \hline 
        \multirow{2}{*}{Hyperparameters} & \multicolumn{2}{c|}{Optimal Values}             \\ \cline{2-3} 
                                 & \multicolumn{1}{c|}{Local Model} & Global Model \\ \hline
        learning\_rate & 4.642e-05 & 2.366e-05 \\
        batch\_size & 256 & 256 \\
        weight\_decay & 1.489e-02 & 1.141e-02 \\ 
        dropout & 0.07572 & 0.05084 \\
        num\_encoder\_layers & 3 & 2 \\
        num\_decoder\_layers & 6 & 3 \\
        d\_model & 96 & 132 \\
        nhead & 3 & 3 \\
        dim\_feedforward & 4096 & 4096 \\
        \hline
    \end{tabular}
    \label{tab:hyperparam_transformer}
\end{table}

\begin{table}[width=\linewidth,pos=h]
    \centering
    \caption{Optimal Hyperparameter Settings for TCN model}
    \begin{tabular}{|l|l|l|}
        \hline 
        \multirow{2}{*}{Hyperparameters} & \multicolumn{2}{c|}{Optimal Values}             \\ \cline{2-3} 
                                 & \multicolumn{1}{c|}{Local Model} & Global Model \\ \hline
        learning\_rate & 0.0436 & 0.0359 \\
        batch\_size & 256 & 256 \\
        weight\_decay & 0.0162 & 0.0268 \\ 
        dropout & 0.4212 & 0.3418 \\
        num\_filters & 4 & 4 \\
        dilation\_base & 5 & 5 \\
        kernel\_size & 4 & 3 \\
        weight\_norm & False & True \\
        \hline
    \end{tabular}
    \label{tab:hyperparam_tcn}
\end{table}

\begin{table}[width=\linewidth,pos=h]
    \centering
    \caption{Optimal Hyperparameter Settings for TFT model}
    \begin{tabular}{|l|l|l|}
        \hline 
        \multirow{2}{*}{Hyperparameters} & \multicolumn{2}{c|}{Optimal Values}             \\ \cline{2-3} 
                                 & \multicolumn{1}{c|}{Local Model} & Global Model \\ \hline
        learning\_rate & 0.0025 & 0.0041 \\
        batch\_size & 256 & 256 \\
        weight\_decay & 1.575e-05 & 1.845e-05 \\ 
        dropout & 0.2666 & 0.1087 \\
        hidden\_continuous\_size & 14 & 17 \\
        hidden\_size & 54 & 54 \\
        lstm\_layers & 3 & 2 \\
        num\_attention\_heads & 2 & 3 \\
        full\_attention & True & False \\
        \hline
    \end{tabular}
    \label{tab:hyperparam_tft}
\end{table}

%%%%%%%%%%%%%%%%%%%%%%%%%%%%%%%%%%%%%%%%%%%%%%%%%%%%%%%%%%%%%%%%
\end{document}